\documentclass{mn2e}
\input{epsf}

\newcommand{\cnu}{\chi^2/\nu}
\newcommand{\blad}[3]{#1^{+#2}_{-#3}}
\newcommand{\g}{$\gamma$}
\newcommand{\msun}{{{\rm M}_{\sun}}}
\newcommand{\xte}{{\it RXTE}}
\newcommand{\pca}{{\it RXTE}/PCA}
\newcommand{\ginga}{{\it Ginga}}

\newcommand{\gro}{{\it CGRO}}

\title[Spectra and variability of GX 339--4]{X-ray and $\bmath{\gamma}$-ray
spectra and variability of the black-hole candidate GX 339--4}

\author[G. Wardzi\'nski et al.]
{Grzegorz Wardzi\'nski,$^1$\thanks{E-mail:
gwar@camk.edu.pl} Andrzej A.~Zdziarski,$^1$ Marek Gierli\'nski,$^{2,3}$ J.~Eric
Grove,$^4$\newauthor Keith Jahoda,$^5$  and W.~Neil Johnson$^4$
\\
$^1$N. Copernicus Astronomical Center, Bartycka 18, 00-716 Warszawa, Poland \\
$^2$University of Durham, Department of Physics, South Road, Durham DH1 3LE,
UK\\
$^3$Astronomical Observatory, Jagiellonian University, Orla 171, 30-244
Krak\'ow, Poland\\
$^4$E. O. Hulburt Center for Space Research, Code 7650, Naval Research
Laboratory, Washington, DC 20375, USA\\
$^5$Laboratory for High Energy Astrophysics, NASA Goddard Space Flight Center,
Greenbelt, MD 20771, USA\\
}

\pagerange{\pageref{firstpage}--\pageref{lastpage}}
\pubyear{2002}

\begin{document}

\maketitle

\label{firstpage}

\begin{abstract} We analyse five observations of the X-ray binary GX 339--4 by
the soft \g-ray OSSE detector on board \gro\/ simultaneous with either
\ginga\/ or \xte\/ observations. The source was bright during four of them, with
the luminosity of $L \sim 10^{37}$ erg s$^{-1}$ and the spectrum typical for
hard states of accreting black holes, and it was in an off state during the
fifth one, with $L \sim 10^{35}$ erg s$^{-1}$. Our broad-band spectral fits show
the mean electron energy of electrons in the Comptonizing plasma decreasing with
increasing luminosity within the hard (bright) state. For the observation with
the best statistics at soft \g-rays, $\sim 1/4$ of energy in the
Comptonizing plasma is probably carried by non-thermal electrons. Then,
considering the efficiency of Comptonized hybrid synchrotron emission allows us
to obtain an upper limit on the strength of the magnetic field in the X-ray
source. Furthermore, this synchrotron emission is capable of producing the
optical spectrum observed in an optically-high state of GX 339--4. In the off
state, the hard X-ray spectrum is consistent with being dominated by
bremsstrahlung. The unusually strong Fe K$\alpha$ line observed by the \pca\
during that state is found not to be intrinsic to the source but to originate
mostly in the Galactic diffuse emission.

\end{abstract}


\section{Introduction}
\label{s:intro}

The Galactic X-ray source GX 339--4 was discovered by the {\it OSO-7\/} 
satellite (Markert et al.\ 1973). It exhibits state transitions typical for 
accreting black holes in binaries; it has been observed in the quiescent, off, 
hard (low), soft (high) and very high X-ray states. Its X-ray spectral analyses 
have been done, e.g., by Ueda, Ebisawa \& Done (1994, hereafter U94), Zdziarski 
et al.\ (1998, hereafter Z98), Wilms et al.\ (1999) and Kong et al.\ (2000). In 
the soft \g-ray domain, OSSE observations have been studied by Grabelsky (1995), 
Z98, and Smith et al.\ (1999), and SIGMA results have been presented by Bouchet 
et al.\ (1993) and Trudolyubov et al.\ (1998). X-ray variability studies have 
been done, e.g., by Maejima et al.\ (1994), Smith \& Liang (1999), Nowak, Wilms 
\& Dove (1999) and Revnivtsev, Gilfanov \& Churazov (2001, hereafter R01).

The source has been also observed in the radio (Fender et al.\ 1999 and 
references therein) and an extended emission from a jet-like elongated structure 
was reported (Fender et al.\ 1997). Optical observations (e.g., Grindlay 1979; 
Motch, Ilovaisky \& Chevalier 1982; Motch et al.\ 1983, hereafter M83; Soria et 
al.\ 1999) have not been able to identify the donor star (see also a discussion 
in Z98). In fact, the observed optical radiation appears to be mostly produced 
in an accretion flow and/or an outflow. This is supported by the decline of the 
optical flux during a transition from a hard to a soft state (M83; Makishima et 
al.\ 1986). Interestingly, this type of transition is also accompanied by a 
decline of the radio flux (Corbel et al.\ 2000).

The presence of an accreting black hole in the system is also suggested by its
Fourier power spectra having the same characteristics at high frequencies as
other black-hole candidates, but not neutron-star systems (Sunyaev \& Revnivtsev
2000). Also, its broad-band energy spectra are remarkably similar to those of a
well established black hole binary, Cyg X-1 (e.g., Gierli\'nski et al.\ 1997,
1999). However, no reliable constraints on the mass of the compact object exist
(see a discussion and references in Z98).

X-ray and $\gamma$-spectra of black holes in their hard states are successfully 
described by models with thermal Comptonization of soft photons in 
optically thin plasmas of a temperature, $T \sim 10^9$ K, and reflection of the 
Comptonization spectrum from an optically thick, cold ($T\sim10^6$ K) medium, 
presumably an accretion disc (Sunyaev \& Tr\"{u}mper 1979, Zdziarski et al.\ 
1995, Gierli\'nski et al.\ 1997). The disc is also likely to serve as the source 
of soft photons undergoing Comptonization. This view is supported by the 
correlation between the spectral index of the Comptonization spectrum and the 
amount of reflection observed in GX 339--4 (U94, R01) and in other accreting 
black holes (Zdziarski, Lubi\'nski \& Smith 1999; Gilfanov, Churazov \& 
Revnivtsev 2000).

The role of magnetic fields remains an open question. They are probably involved
in transfer and generation of energy either via viscosity mechanisms or
reconnection processes. As a consequence, synchrotron radiation should be
present and it may serve as a source of soft photons undergoing further
Comptonization. Though for thermal plasmas this process appears to be typically
negligible (Wardzi\'nski \& Zdziarski 2000, hereafter WZ00), the presence of a
weak non-thermal tail strongly amplifies synchrotron emission (Wardzi\'nski \&
Zdziarski 2001, hereafter WZ01). Then, the synchrotron process can become a very
effective source of soft photons, comparable to or stronger than the cold disc
emission.

Consequently, detection of a non-thermal electron tail via a high-energy tail in
a Comptonization spectrum can put constraints on the strength of the magnetic
field. In particular, observations of Cygnus X-1 in its hard state suggest the
presence of such a tail (McConnell et al.\ 2000, 2002). For this case, WZ01
obtained an upper limit on the magnetic field strength in the accretion flow and
found constraints on the geometry.

Here, we study those (and some other) issues in the context of GX 339--4. We 
analyze five \ginga\/ and \xte\/  observations of GX 339--4 which were 
simultaneous (or nearly simultaneous) with $\gamma$-ray observations by the OSSE 
detector onboard \gro. This provides us with data in the useful range from a few 
keV to a few hundred keV. The data reduction procedures and models we use are 
described in Sections \ref{s:data} and \ref{s:models}, respectively. We present 
the results of spectral analyses in Sections \ref{hard} and \ref{off}. 
Consequences of the possible presence of a non-thermal electrons for accretion 
flow models are discussed  in Section \ref{s:nth}. We investigate the connection 
between the X-ray and optical data in the framework of synchrotron self-Compton 
models in Section \ref{s:optical}, while X-ray variability is examined in 
Section \ref{s:variability}. Our results are summarized in Section 
\ref{s:summary}.

\section{Selection and reduction of observational data}
\label{s:data}

The data used in this work are listed in Table 1. The first two data sets 
are identical with those of Z98. They consist of two \ginga\/ data sets from 
1991 September 11 and 12, with the usable energy range of 1.2--30 keV, and 
simultaneous OSSE data in the range from 50 keV up to several hundred keV. A 1 
per cent systematic error is added in quadrature to the statistical error in 
each \ginga\/ channel (as in U94). All the OSSE data used here include 
energy-dependent systematic errors estimated from the uncertainties in the 
low-energy calibration and response of the detectors using both in-orbit and 
prelaunch calibration data. They decrease from $\sim 3$ per cent at 50 keV to 
$\sim 0.3$ per cent at $\ga 150$ keV.

\begin{table*}
\centering
\begin{minipage}{175mm}
\caption{A summary of the observational data. The background-subtracted count
rates  for the LAC, PCA, HEXTE and OSSE instruments are for the 1.2--29, 3--30,
30--100 and 50--1000 keV energy ranges, respectively.}
\begin{tabular}{lccccccc}
\hline
No.&Satellite&Date&Observation ID/&Instrument&Exposure\footnote{For HEXTE, the detector live time (i.e.\ dead-time corrected) is given.}&Count rate &References\\
&&&Viewing Period&&[s]&[s$^{-1}$]&\\\\
1&\ginga&1991 Sept.\ 11&91091109421&LAC&$2.0\times 10^3$&1720&U94; Z98
\\
 &\gro&1991 Sept.\ 11&VP 9&OSSE&$4.8\times 10^3$&15&Grabelsky et al.\ (1995)\\[0.4em]
2&\ginga&1991 Sept.\ 12&91091109421&LAC&$3.8\times 10^2$&1700&U94; Z98
\\
 &\gro&1991 Sept.\ 12&VP 9&OSSE&$2.5\times 10^3$&15&Grabelsky et al.\ (1995)\\[0.4em]
3&\xte&1996 Jul.\ 26&P10420-01-01-00&PCA&$5.2\times 10^3$&271&B\"ottcher et al.\ (1998); \\
&&&&HEXTE A&$1.7\times10^3$&16&Smith \& Liang (1999);\\
&&&&HEXTE B&$1.7\times10^3$&13\\
&\gro&1996 Jul.\ 9--23&VP 524&OSSE&$5.4 \times 10^5$&6.5& Smith et al.\ (1999)\\[0.4em]
4&\xte&1997 Feb.\ 14 &P20183-01-02-00, 01&PCA&$2.1\times 10^4$&217&R01\\
&&&&HEXTE A&$6.5 \times 10^3$&13&\\
&&&&HEXTE B&$3.6 \times 10^3$&11&\\
&\gro&1997 Feb.\ 11--19&VP 614.5&OSSE&$3.4\times 10^5$&7.0&\\[0.4em]
5&\xte&1999 Jul.\ 7&P40108-03-01-00&PCA&$1.2\times 10^4$&1.5& Feng et al.\ (2001)\\
&\gro&1999 Jul.\ 6--13&VP 821&OSSE&$1.1\times 10^5$&0.2&\\
\hline
\end{tabular}
\label{t:data}
\end{minipage}
\end{table*}

The third data set consists of a relatively short \xte\/  observation on 1996 
July 26, which was performed shortly after a two-week long OSSE observation of 
1996 July 9--23. For these \xte\/  data (and other ones used here), we used data 
from the units 0 and 1 of the PCA detector, and used only layer 1, which 
provides the highest signal-to-noise ratio. This choice is favoured by Wilson \& 
Done (2001) based on their study of Crab spectra from the PCA. Also following 
them, we added a systematic error of 0.5 per cent to the PCA data. The \xte\/ 
data were reduced with the {\sc lheasoft} package v.\ 5.0.4, response matrix v.\ 
7.10. Standard data selectrion criteria were applied, namely the Earth elevation 
angle $>10^\circ$, pointing offset $<0.02^\circ$, the time since the peak of the 
last SAA passage greater than 30 min and the electron contamination  $<0.1$. We 
used the Sky\_VLE background model for bright sources dated 1999 September. We constrain 
the energy range to a conservative range of 3--30 keV. We also used data from 
the clusters A and B of the HEXTE detector in the 30--100 keV range.

The fourth data set consists of an \xte\/  observation on 1997 February 14 and
an OSSE observation on 1997 February 11--19. This data set has the best
statistics among all ones used here.

Finally, the fifth data set includes an \xte\/  observation on 1999 July 7 and a
two-week long OSSE observation of 1999 July 6--13. In the PCA energy range, the
source was by two orders of magnitude fainter than in the other data sets. Thus,
in order to obtain the best signal to noise ratio, in the PCA data analysis we
used the layer-1 data from all the detector units available during this
observation, i.e., 0, 2 and 3. Here we employed the background model for faint sources from 2002 February. The HEXTE data were not used because of very
large statistical errors after background subtraction.

\section{Spectral models}
\label{s:models}

Fits below are performed with the {\sc xspec} package (Arnaud 1996) v.\
11. The confidence ranges of each model parameter are given for a 90 per cent
confidence interval, i.e., $\Delta \chi^2=2.71$ (e.g.\ Press et al.\ 1992). On
the other hand, the plotted vertical error bars are 1-$\sigma$, the upper
limits, 2-$\sigma$, and the plotted spectral data are rebinned for clarity of
the display.  Spectra from various detectors are allowed to have free relative
normalization.

In most cases, the intrinsic continuum is modelled by Comptonization of
blackbody photons in a hybrid plasma with the {\sc compps} model of Poutanen \&
Svensson (1996). Compton reflection of the continuum from an ionized, optically
thick medium (presumably a cold accretion disc) is also taken into account using
the inclination-dependent Green's functions of Magdziarz \& Zdziarski (1995).

In the case of a purely thermal plasma, the Comptonization spectrum is 
characterized by the Thomson optical depth, $\tau$, the electron temperature 
(which we express in energy units, $kT$, where $k$ is Boltzmann constant), and 
the temperature of the blackbody seed photons, $kT_{\rm seed}$. We also consider 
a hybrid electron distribution, which consists of a Maxwellian up to a electron 
Lorentz factor, $\gamma_{\rm nth}$, and a power-law of an energy index, $p$, 
above it. Given these parameters, we calculate the ratio, $\delta$, of the 
energy in the non-thermal electrons to that in the thermal ones. In order to 
quantify the hardness of a given Comptonization spectrum, we also calculate the 
4--20 keV power-law photon index, $\Gamma$, which we obtain by fitting data 
simulated with the model of (only) Comptonization.

The reflection is characterized by the solid angle subtended by the reflecting 
medium, $\Omega$, its ionization parameter, $\xi$, and the Fe abundance relative 
to the solar one (Anders \& Ebihara 1982), $A_{\rm Fe}$. Here, $\xi\equiv L_{\rm 
ion}/n R^2$, $L_{\rm ion}$ is defined as the 5 eV--20 keV luminosity in a power 
law spectrum and $n$ is the density of the reflector located at distance $R$ 
from the illuminating source. We use here the ionization model by  Done et al.\ 
(1992), which is not applicable to very highly ionized Fe (Ballantyne, Ross \& 
Fabian 2001), but sufficiently accurate at low/moderate ionization, such as that
found in our data. The reflector temperature is kept at $10^6$ K, which is about 
the highest temperature consistent with the model of ionization equilibrium of 
Done et al.\ (1992). Compton reflection is accompanied by a fluorescent Fe 
K$\alpha$ line at a rest-frame energy, $E_{\rm Fe}$, and an equivalent width, 
$W_{\rm Fe}$. Both the line and reflection are relativistically broadened in an 
accretion disc in the Schwarzschild metric with the emissivity equal to that of 
a thin disc, extending from an outer radius, $r_{\rm out}$ (assumed to equal 
$10^3$), to an inner one, $r_{\rm in}\geq 6$, where $r$ is in the units of 
$GM/c^2$ (Fabian et al.\ 1989). We note here that the values of $kT$, 
$\gamma_{\rm nth}$, and $p$ can be constrained mostly by the OSSE data whereas 
$\tau$ (for given $kT$) and the reflection/line parameters are determined mostly 
by either \ginga\/ or \xte\/ data.

We assume the column density to the source of $N_{\rm H}= 6\times 10^{21}$ 
cm$^{-1}$. This value was derived from $E(B-V)$ measurements and is consistent 
with results of other methods, see Z98 and references therein. We assume the 
distance to the source of $d=4$ kpc (which agrees with both kinematic and 
extinction measurements, see Z98). In calculations of reflection and 
relativistic broadening, we assume an inclination of $i=45^\circ$, and we also 
assume $M=10\msun$, but we stress that currently there are no constraints on 
these quantities.

\section{Hard state observations}
\label{hard}

The source flux (extrapolated to the range of 1--1000 keV) in the data sets from
1 to 4 is $F \sim 10^{-9}$ erg cm$^{-2}$ s$^{-1}$, which corresponds to the
luminosity of $L \sim (1$--$3) \times 10^{37}$ erg s$^{-1}$ and $\sim 1$ per
cent of the Eddington luminosity (at $d=4$ kpc and $M=10\msun$ assumed here).

\subsection{The \textbfit{G\lowercase{inga}}-OSSE data}
\label{ss:ginga}

Results of thermal-Comptonization fits to the \ginga-OSSE data sets 1 and 2 are
given in rows 1 and 2 of Table \ref{t:fits}, respectively.  The former has much
better statistics and we discuss it in more detail, but results for the set 2
are very similar. As it was shown by Z98 (see their section 5.1), spectral fits
to the \ginga-OSSE data do not constrain the geometry of the Comptonizing plasma
and the seed-photon source. Thus, we assume here a simple geometry of a
spherical source with seed photons emitted uniformly. We also account for the
soft excess present in the data by a blackbody component at the same temperature
as that of the seed photons.

We obtain a very good fit to the data set 1, with $\cnu =46/81$. The value of 
$\cnu$ much smaller than unity results from relatively large systematic errors 
added to the \ginga\/ data, see Section \ref{s:data}. The fit components, as 
well as comparison of this spectrum with spectra from data set 3 and 4 (analyzed 
in Section \ref{s:xte} below) are shown in Fig. \ref{f:spectra}. Fig. 
\ref{f:chi} shows the quality of our fits to the data sets 1--4.  
The parameters of the fit to data set 1 are very similar to those obtained by 
Z98 and indicative of a typical hard-state spectrum of black-hole binaries. The 
main difference with respect to the results of Z98 in their table 2 is the 
plasma temperature, which we fit here at $kT = \blad{46}{6}{4}$ keV, whereas Z98 
obtained $kT = \blad{57}{7}{5}$ keV. This difference is due to the model of 
Comptonization used here being more accurate than that used by Z98.

The fitted relative normalization of OSSE to \ginga\/ is close to unity, $\blad{0.91}{0.07}{0.07}$
and $\blad{0.90}{0.1}{0.1}$ for the sets 1 and 2, respectively. Hereafter, we give the luminosity
normalized to the OSSE data. Such a choice is appropriate for investigating a
connection between the luminosity and $kT$ (see Section \ref{T_L})
since both are determined mostly by the OSSE data.

\begin{table*}
\caption{Main parameters of the fits in Section \ref{s:models}. The numbers
correspond to the data sets in Table \protect\ref{t:data} and a and b denote the
thermal and hybrid fits, respectively, to the set 4.  $kT$, $L$(1--1000 keV),
$\xi$, $E_{\rm Fe}$, and $W_{\rm Fe}$ are is units of keV,  $10^{37}$ erg
s$^{-1}$,  erg s$^{-1}$ cm,  keV and eV, respectively. }
\begin{tabular}{lccccccccccccc}
\hline
Fit&$kT$&$L$&$\tau$&$\Gamma$&$\Omega/2\upi$&$\xi$&$A_{\rm Fe}$&$kT_{\rm seed}$
&$E_{\rm Fe}$&$W_{\rm Fe}$&$r_{\rm in}$&$\cnu$\\
\\
1&$\blad{46}{6}{4}$&$2.5$&$\blad{2.5}{0.3}{0.3}$&$1.75$&$\blad{0.36}{0.07}{0.07}
$&$\blad{130}{110}{100}$&$\blad{2.9}{1.7}{0.9}$&$\blad{0.25}{0.02}{0.03}$
&$\blad{6.58}{0.32}{0.32}$&$\blad{70}{43}{34}$&$\blad{36}{\infty}{30}$&46/81\\
2&$\blad{43}{8}{6}$&$2.3$&$\blad{2.7}{0.4}{0.3}$&$1.73$&$\blad{0.21}{0.09}{0.09}
$&$\blad{340}{1050}{260}$&$\blad{1.6}{3.6}{0.9}$&$\blad{0.27}{0.03}{0.03}$
&$\blad{6.58}{0.30}{0.32}$&$\blad{61}{34}{33}$&$\blad{185}{\infty}{179}$&63/81\\
3&$\blad{58}{3}{3}$&$0.88$&$\blad{1.9}{0.1}{0.1}$&$1.79$&$\blad{0.40}{0.06}{0.06
}$&$\blad{110}{70}{40}$&$\blad{3.6}{1.1}{1.8}$&$\blad{0.21}{0.06}{0.10}$
&$\blad{6.28}{0.14}{0.14}$&$\blad{88}{17}{18}$&$6^{+65}$&159/130\\
4a&$\blad{76}{5}{6}$&$1.2$&$\blad{1.5}{0.1}{0.1}$&$1.79$&$\blad{0.39}{0.04}{0.03
}$&$\blad{90}{30}{30}$&$\blad{2.8}{0.8}{0.6}$&$\blad{0.16}{0.05}{0.15}$
&$\blad{6.31}{0.12}{0.13}$&$\blad{59}{11}{12}$&$\blad{33}{53}{27}$&143/148\\
4b&$\blad{46}{11}{8}$&$1.2$&$\blad{2.2}{0.4}{0.3}$&$1.76$&$\blad{0.27}{0.05}{0.
06}$&$\blad{330}{460}{170}$&$\blad{2.3}{0.8}{0.7}$&$\blad{0.33}{0.04}{0.06}$
&$\blad{6.23}{0.11}{0.10}$&$\blad{105}{11}{13}$&$6^{+10}$&133/147\\
\hline
\end{tabular}
\label{t:fits}
\end{table*}

\begin{figure}
\begin{center}
\noindent\epsfxsize=8.6cm\epsfbox{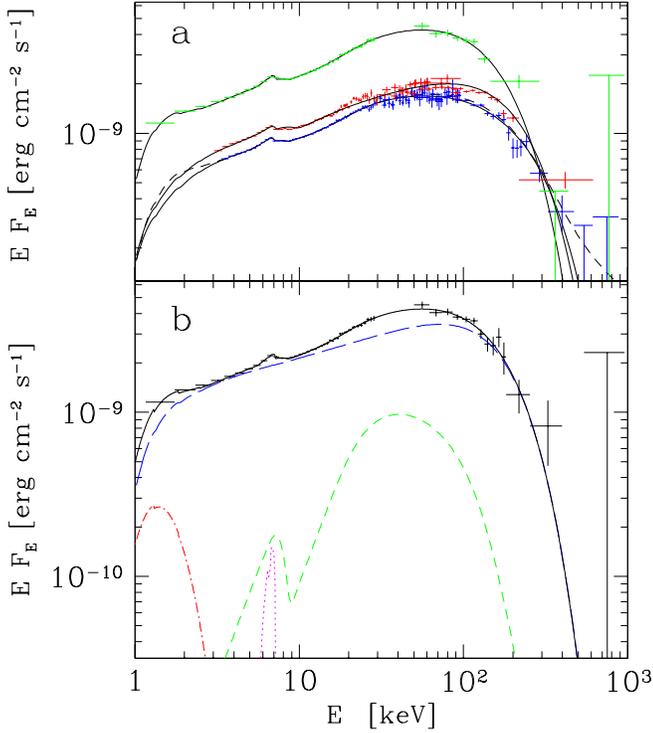}
\end{center}
\caption{(a) Broad-band spectra of GX 339--4. The data sets 1, 3 and 4 (with
decreasing flux) are shown in the green, red and blue error bars, respectively.
The black solid curves give the best thermal fits to the data, and the dashed
curve gives the best non-thermal fit to the set 4. (b) The spectral components
of the model of the set 1 shown separately. Thermal Comptonization, the soft
excess, Compton reflection and the Fe K line are shown by the
long-dashed blue, dot-dashed red, short-dashed green and dotted magenta curves, respectively. The solid black line gives the total spectrum. Absorption by the interstellar medium is seen below a few keV.
\label{f:spectra}
}
\end{figure}

\begin{figure*}
\begin{center}
\noindent\epsfxsize=8.45cm\epsfbox{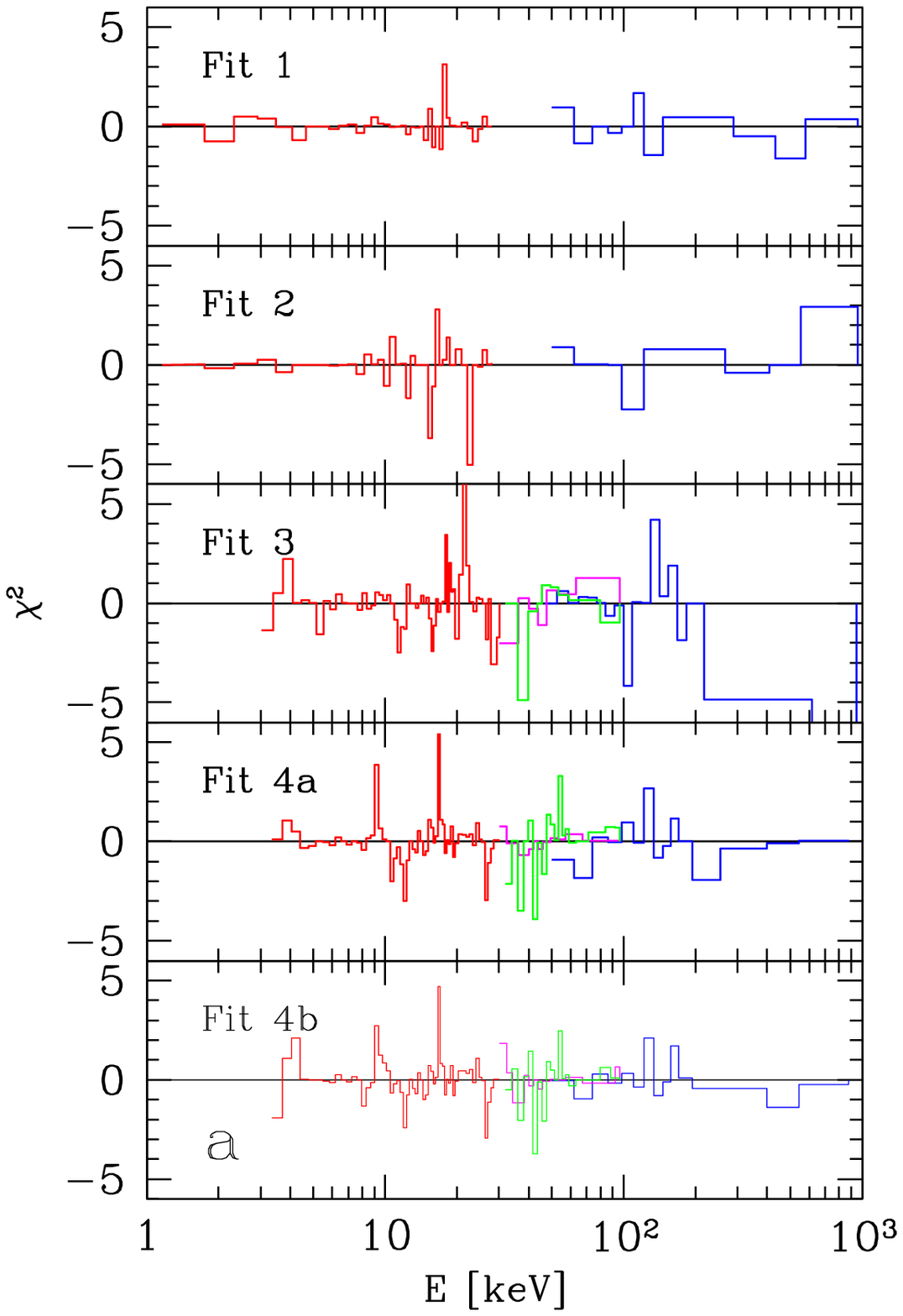}\noindent\epsfxsize=8.6cm\epsfbox{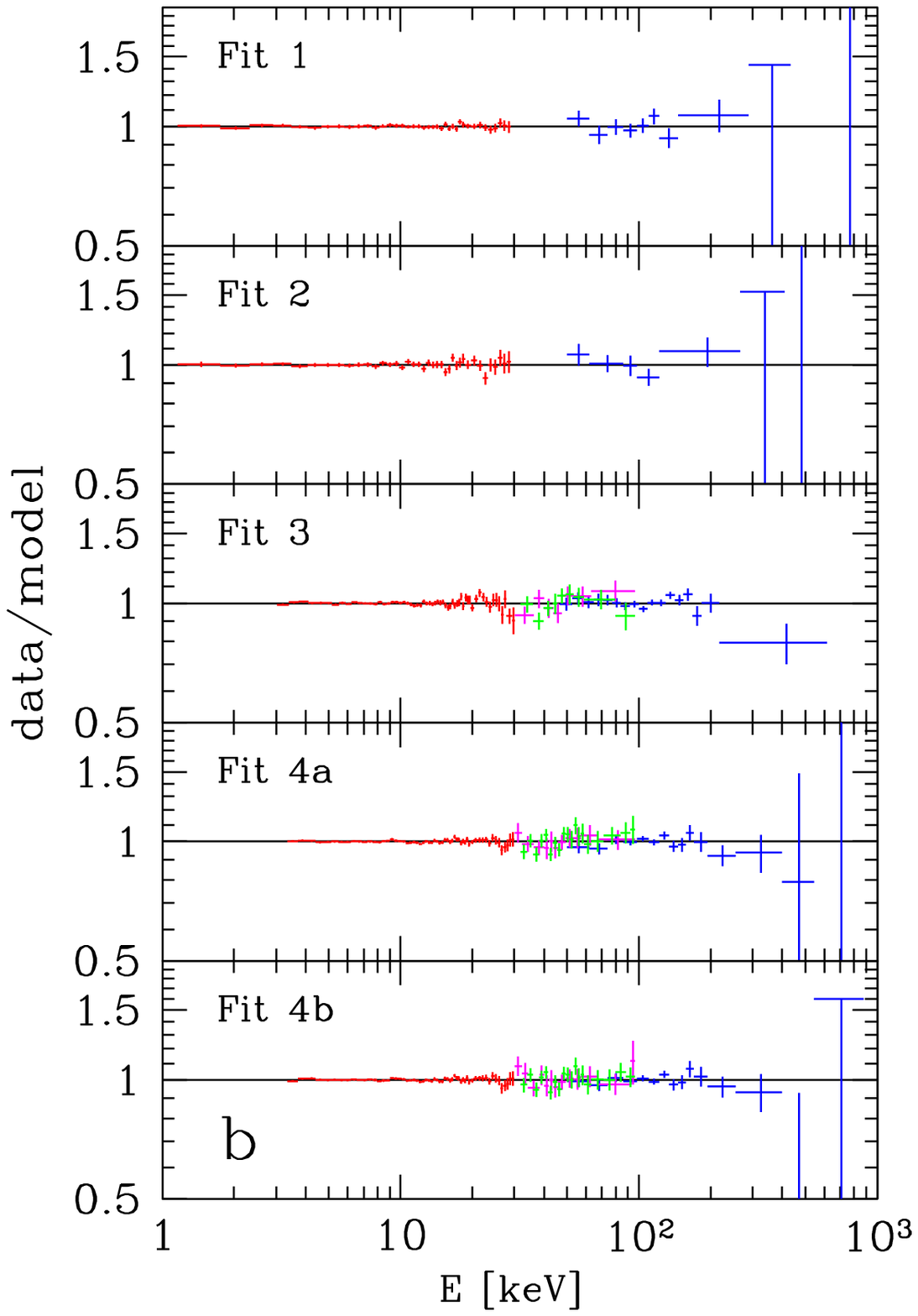}
\end{center}
\caption{(a) The contribution to the total $\chi^2$ from data bins multiplied by the sign of (data-model) and (b) data-to-model ratios for fits presented in Table \protect\ref{t:fits}. Red lines represent {\it Ginga}\/ or PCA data, while green, magenta and blue lines correspond to the HEXTE A, HEXTE B and OSSE data, respectively.}
\label{f:chi}
\end{figure*}

In non-homogenous geometries (such as a corona above a disc), an anisotropy 
break appears at the peak energy of the second-order scattering for a high 
enough $kT$, with deficiency of photons below the break with respect to the 
extrapolation of the main power law (e.g., Poutanen \& Svensson 1996). In 
general, the form or the absence of this break may be then used to constrain the 
geometry. However, the energy range where it may appear in the case of GX 339--4 
is similar to that of the soft excess. Consequently, fits assuming different 
geometries and including the soft excess give very similar statistics [see also 
discussions in Z98 and Nowak, Wilms \& Dove (2002)]. The presence of the soft 
excess itself is very significant statistically, e.g., $\Delta \chi^2=+29$ for a 
fit to the set 1 in spherical geometry and without the soft excess.

We also note that our adopted simple model of the soft excess as a blackbody is 
probably not realistic. In particular, the fitted $kT_{\rm seed}\sim 0.3$ keV 
is about twice higher than that of Cyg X-1 in the hard state (Ebisawa et al.\ 
1996). In Cyg X-1, the soft excess is consistent with the presence of both blackbody and its Comptonization in a plasma cloud separate from the main Comptoning source
(Di Salvo et al.\ 2001; Frontera et al.\ 2001a). 
This results in the onset of the excess at $\sim 3$ keV (similar to GX 339--4), 
but with the excess $\sim 1$--3 keV emission being mostly due to Comptonization, 
not blackbody. Consequently, we also have fitted the soft excess by a second 
thermal Comptonization component (in addition to the main Comptonization 
component responsible for the emission at $\ga 3$ keV). We find the excess can 
be fitted by emission of either an optically-thick plasma with $\tau=6.1$ and 
$kT=1.4$ keV (using {\sc comptt}, Titarchuk 1994), or optically-thin one with 
$\tau=0.43$, $kT=20$ keV (using {\sc compps}). Both models yield $\cnu =45/79$, 
and are analogous to the models of the soft excess in Cyg X-1 by Di Salvo et 
al.\ (2001) and Frontera et al.\ (2001a), respectively. We conclude that the 
data do not allow us to distinguish between different soft excess models.

As far as the high-energy end of the measured spectrum is concerned, an
important issue is the possible presence of a high-energy tail in the electron
distribution. We have found that none of the OSSE data considered in this work
have enough statistics at high energies to constrain the electron power law
index, $p$. We have thus chosen $p=4$, which is close to the values
measured in Cyg X-1, see Gierli\'nski et al.\ (1999), and McConnell et al.\
(2000, 2002), and it was used in the theoretical study of WZ01.

We find that the \ginga-OSSE data are consistent with the presence of a 
high-energy electron tail, but they do not constrain its amplitude. E.g., the 
non-thermal fraction, $\delta$, for the data set 1 is $0.03_{-0.03}^{+0.31}$. 
The presence of a tail is indicated by the result of Z98, who obtained $\Delta 
\chi^2=-4$ when allowing for the presence of non-thermal electrons in their fit 
to the OSSE data from 1991 September 5--12. However, an alternative explanation 
of the departure of the spectrum for the thermal one can be a distribution of 
the plasma parameters, in particular $kT$, either in time or in space.

\subsection{The \textbfit{RXTE}-OSSE data}
\label{s:xte}

Since the usable PCA data extend down only to 3 keV (compared to 1.2 keV for the 
\ginga\/ data), the soft excess cannot be constrained by them at all. Thus, we 
neglect its presence for all \xte-OSSE data. Then, the thermal model applied to 
the data set 4 (fit 4a in Table \ref{t:fits}) yields an acceptable $\cnu= 
143/148$. The normalization of the models to the HEXTE A and B clusters relative 
to the PCA data is $\blad{0.88}{0.03}{0.03}$ and $\blad{0.85}{0.03}{0.03}$, respectively. On the other hand, the relative 
normalization of models to the OSSE and PCA data was almost unity, $\blad{0.98}{0.03}{0.03}$, in 
spite of the observing time being larger for the OSSE. The results of our fits 
in the X-ray regime are similar to those of Wilms et al.\ (1999), who analyzed 
an \xte\/ observation that took place three days after ours.

Our data set 4 has the best quality of all those considered here, and allows us
to study the presence of a non-thermal electron tail in the source. We have
found it improves the fit (4b in Table \ref{t:fits}) very significantly, to
$\cnu=133/147$, with the probability of the fit improvement being by chance of
only $10^{-3}$ (using the F-test, Bevington \& Robinson 1992). The parameters of the
hybrid electron distribution are $\gamma_{\rm nth}=\blad{1.59}{0.24}{0.14}$ and
$kT= \blad{46}{11}{9}$ keV, which corresponds to $\delta=
\blad{0.34}{0.22}{0.17}$ ($\delta$ increases with decreasing either $kT$ or
$\gamma_{\rm nth}$). The normalizations of the models to the HEXTE A and B clusters and to the OSSE detector, relative to the PCA data, were $\blad{0.86}{0.03}{0.03}$, $\blad{0.84}{0.03}{0.03}$ and $\blad{0.92}{0.04}{0.04}$, respectively.

We caution, however, that the OSSE observation was much longer than the \xte\/
one, and thus it is possible that the modification of the shape of the OSSE
spectrum was due not to non-thermal electrons, but resulted from variations of
the plasma parameters in time or in space. We have checked these possibilities.
First, we allowed for different values of $\tau$ in thermal fits to the \xte\/
and OSSE data. This, however, did not change the fit quality, $\cnu = 142/147$,
and the difference between the two values of $\tau$ was only 0.1. We then fit
two {\sc compps} models allowing for different values of $\tau$ and $kT$ and
keeping all other model parameters the same. This significantly improves the
fit compared to the previous thermal model, resulting in $\cnu=135/145$, only
slightly worse than the hybrid fit. The fitted plasma parameters are $kT=61$
keV, $\tau=2.1$, and $kT=83$ keV, $\tau=1.1$ for the two components. Indeed,
plasmas in an accretion flow are expected, on average, to increase its
temperature with decreasing optical depth as a result of less effective Compton
cooling, in agreement with our fit results. Thus, the data set 4 is as
consistent with emission from regions of different $kT$, $\tau$ as with the
presence of a non-thermal tail in a uniform electron population. The latter
interpretation may be favoured by the similarity of our result to those of
McConnell et al. (2000, 2002) for Cyg X-1.

On the other hand, either thermal or hybrid model applied to the data set 3
yields unacceptable fits, e.g., $\cnu = 202/131$ in the thermal case. However,
both the OSSE lightcurve and that from the All-Sky Monitor on board \xte\/
(available on the public database at xte.mit.edu/ASM\_lc.html) have shown that
the source was significantly variable on timescale of days during the
observation. In particular, the OSSE spectrum underwent a significant softening
and brightening during the two-week observation (Smith et al.\ 1999). The
softening implies either $kT$ or $\tau$ of the plasma decreasing with time, to
that during the \xte\/ observation (three days after the OSSE one). We have
accounted for this effect by allowing for different values of $\tau$ and the
same $kT$ (since the \xte\/ data alone constrain it only weakly) in the fits to
the \xte\/ and the OSSE data. We find the best-fit $\tau$ for the OSSE data of
2.4, indeed larger than the value for the \xte\/ data (Table \ref{t:fits}).
Then, the model becomes statistically acceptable, with $\cnu=159/130$. The
relative normalization of the OSSE and PCA data corresponds to the ratio of the
respective model luminosities of 0.67, while the normalizations of the HEXTE A and B data, relative to the PCA data, were $\blad{0.89}{0.04}{0.04}$ and $\blad{0.82}{0.04}{0.04}$, respectively.

Our results differ significantly from those of Smith et al.\ (1999) and
B\"ottcher et al.\ (1999). They analyzed the same data and concluded that the
\xte-OSSE spectrum could be modelled without reflection. In order to reproduce
the hardening at $\sim 10$ keV (characteristic to reflection), they introduced
two different power-law models intersecting at $\sim 10$ keV. We consider this
approach (analogous to that of Dove et al.\ 1998 applied to Cyg X-1) to
represent an only phenomenological (i.e., unphysical) description of the
spectrum. The hardening at $\sim 10$ keV is, at very high probability, indeed
due to reflection (and not due to a fortitous intersection of two power law
components just at 10 keV), as it is present in many other spectra of GX 339--4
as well as in many other Galactic and extragalactic accreting black-hole sources
(e.g., Gilfanov et al.\ 2000; Nandra \& Pounds 1994). In fact, the models of
Smith et al.\ (1999) and B\"ottcher et al.\ (1999) do include an Fe K$\alpha$
line and an Fe K edge, which features naturally appear in the presence of
Compton reflection.

In the X-ray data (from both \ginga\/ and \xte), the reflector was found to be 
moderately ionized, with $\xi \sim 10^2$ erg s$^{-1}$ cm (see discussion of 
physical implications in Z98). In the \ginga\/ data, $E_{\rm Fe}$ was $>6.4$ keV 
and consistent with modest ionization. On the other hand, it was $<6.4$ keV in 
the \xte\/ data, where disc ionization was also high. Such an effect was also 
found in other \xte\/ observations of GX 339--4 (Wilms et al.\ 1999; Feng et 
al.\ 2001), which may indicate a calibration problem. The equivalent width of 
the line, $W_{\rm Fe} \sim 70$ eV or so, is roughly consistent with the observed 
reflection (George \& Fabian 1991; \.Zycki  \& Czerny  1994). We also see in 
Table \ref{t:fits} that relativistic smearing of the reprocessing features 
cannot be constrained by the data. Only the \xte\/ data appear to constrain 
$r_{\rm in}$ to $\la 10^2$.

\subsection{A possible relation between the high-energy cutoff and luminosity}
\label{T_L}

The four bright-state observations of GX 339--4 cover a factor of $\sim 3$ in 
the luminosity, $L$. On the other hand, all the spectra have very similar values 
of the X-ray slope, $\Gamma\simeq 1.75$, as seen in Table \ref{t:fits}. An 
interesting issue is then a possible correlation between $L$ and the high-energy 
cutoff in the spectrum. Such a correlation is generally predicted by theories of 
accretion flows, as a higher accretion rate, $\dot M$, usually yields higher 
both $L$ and $\tau$. If the X-ray slope is constant, the electron temperature 
(determining the cutoff) should then decrease, an effect due to an increased 
efficiency of cooling per electron.

\begin{figure}
\begin{center}
\noindent\epsfxsize=6.8cm\epsfbox{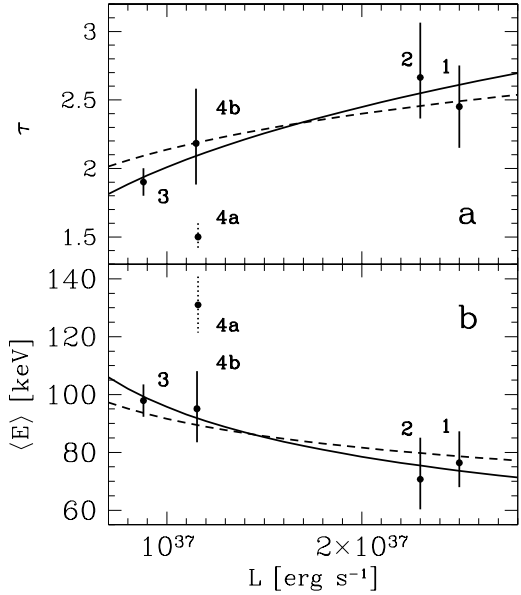}
\end{center}
\caption{Relation between the luminosity and (a) the Thomson optical depth,
and (b) the average electron energy. Labels correspond to the fit numbers from
Table \protect\ref{t:fits}, and the dotted line denotes the purely thermal model
to the data set 4, which is statistically less likely than the hybrid model.
Solid and dashed curves show the fitted dependences on $L$ in the cases of
advection and cooling dominated flows, respectively (see Section \protect
\ref{T_L}).}
\label{f:correl}
\end{figure}

A complication for our data arises due to the possible presence of non-thermal
electrons. A purely thermal fit to a hybrid plasma will overestimate the actual
$kT$, as can be seen by comparing the results of the fits 4a and 4b in Table
\ref{t:fits}. On the other hand, both thermal and non-thermal electrons Compton
scatter photons, and the average electron energy, $\langle E\rangle$, is then
the crucial factor determining the low-energy slope of the Comptonization
spectrum and the position of the high-energy cutoff. These effects are taken
into account in the {\sc compps} model. We can then look directly for a
correlation between $L$ and $\tau$ using fits with that model. Fig.\
\ref{f:correl}a shows that $\tau$ indeed increases almost monotonically with
increasing $L$ provided the hybrid fit is used for the data set 4, supporting
the presence of a non-thermal tail in that observation. Similarly, $\langle
E\rangle$ shows a decreasing trend with increasing $L$, as shown in Fig.\
\ref{f:correl}b. (See also Zdziarski 2000 for a comparison of the spectra from
the sets 1 and 4.)

Zdziarski (1998) considered a simple hot accretion model parametrized by the
Compton parameter, $y$ ($=4 \tau kT /m_{\rm e} c^2$ in the thermal case, and
$\simeq 1$ for our data). The Compton parameter determines in turn the X-ray
slope, $\Gamma$, which is constant among the 4 data sets. For constant $y$,
$\tau\propto L^{2/7}$ and $L^{1/6}$ in the advection and cooling dominated
cases, respectively (Zdziarski 1998). The two dependences are shown in Fig.\
\ref{f:correl}a by the solid and dashed curves, respectively. We see that the
former appears to fit the data slightly better, but the result is not conclusive. The corresponding relations for $\langle
E\rangle$ are shown in Fig.\ \ref{f:correl}b, where we assumed  $y\propto
\langle E\rangle \tau$.

Physically, a constant $y$ (or $\Gamma$) corresponds to a constant ratio of the
Comptonization luminosity to that in the seed photons (e.g., Beloborodov 1999b),
which can be due to a constant geometry of the flow. We also note that the
values of $L$ in our data are close to the maximum in a hot accretion flow with
advection and Comptonization cooling, $L \approx 10^{37} y^{3/5} (\alpha_{\rm
v}/0.1)^{7/5} (M/10 \msun)$ (Zdziarski 1998), where $\alpha_{\rm v}$ is the
viscosity parameter.

\section{The off-state spectrum}
\label{off}

During the fifth observation (Table \ref{t:data}), GX 339--4 was by 2 orders of
magnitude fainter than during the hard-state observations. At this flux level,
GX 339--4 has been  classified as being in the off state, see, e.g., Kong et
al.\ (2000). We note here that the object also enters even more quiescent
states, e.g., during an observation of Asai et al.\ (1998), when an upper limit
on the source flux was about two orders of magnitude below the flux of our data.

We first fit the data with the same thermal model as for the sets 3--4, in which
we assume the relative normalization of the OSSE data to the PCA ones of unity.
This provides a statistically acceptable model, with $\cnu=46/64$, unconstrained
reflection, and a broad range of the allowed electron temperature, $kT =
\blad{170}{230}{150}$ keV. However, the spectrum shows a distinct hardening at
$\sim 7$ keV, which in turn requires either a very high temperature of the seed
photons, $kT_{\rm seed} \sim 1$ keV, or very strong Compton reflection,
$\Omega/2\upi \gg 1$. The former is rather unlikely at the observed low
luminosity level, $L=8 \times 10^{34}$ erg s$^{-1}$, and the latter has never
been observed in any other black-hole binary.  We have performed extensive tests
checking also the possibility that the hardening is due to the PCA background
subtraction uncertainties. We have concluded it to be unlikely and hereafter
consider the hardening to be intrinsic to the source, but still cannot rule out
completely a significant effect of the background subtraction.

\begin{figure}
\begin{center}
\noindent\epsfxsize=7.8cm\epsfbox{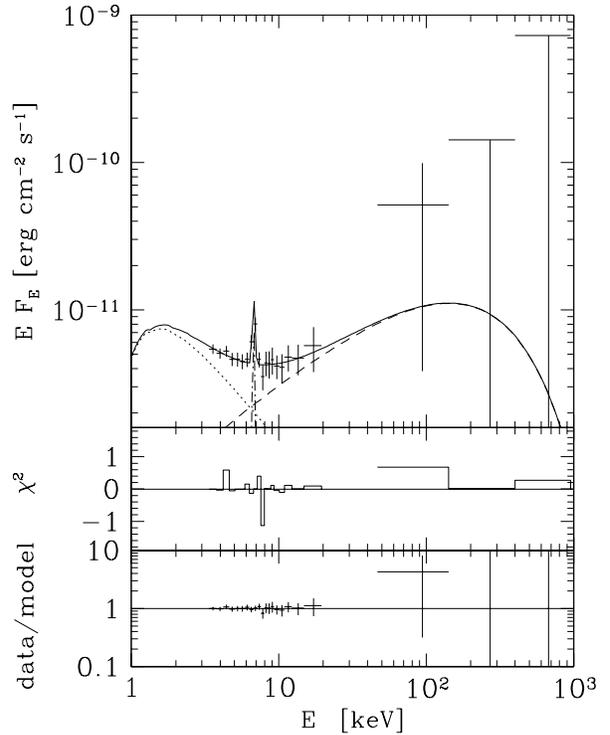}
\end{center}
\caption{The upper panel shows the off-state spectrum (error bars) from the fifth observation, modelled by thermal bremsstrahlung (dashed curve), power-law Comptonization at low energies (dotted curve), and an Fe K line (dot-dashed curve). The solid curve gives the sum. Lower panels give the contribution from data bins to the total $\chi^2$ and the data to model ratio.}
\label{f:off}
\end{figure}

On the other hand, it is possible that spectrum below several keV is the 
standard Comptonization power law, and the hardening at higher energies is due 
to thermal bremsstrahlung. That process is indeed expected to dominate the 
high-energy spectrum at very sub-Eddington accretion rates, see, e.g., Narayan 
\& Yi (1995), Esin, McClintock \& Narayan (1997), Zdziarski (1998). We model the 
spectrum by the sum of a power law (corresponding to Comptonization at low 
energies), thermal bremsstrahlung and a Gaussian Fe K$\alpha$ line. This yields 
a good fit, with $\cnu = 46/66$. The slope of the low-energy power law is 
$\Gamma = \blad{3.1}{1.5}{0.7}$ whereas the bremsstrahlung temperature is 
unconstrained. We have assumed it to equal 200 keV (e.g., Esin et al.\ 1997), 
which corresponds to $L\sim 10^{35}$ erg s$^{-1}$, and, e.g., $\tau \approx 
0.35$ and the source radius of $R\approx 10^9$, with the bremsstrahlung 
luminosity scaling approximately as $T^{1/2} \tau^2 R$. The data and the model 
spectrum are shown in Fig.\ \ref{f:off}.

An important issue here is the origin of the strong Fe K$\alpha$ line seen in
the data. In the model with bremsstrahlung, the line is narrow at $E_{\rm
Fe}=\blad{6.8}{0.3}{0.4}$ keV with the photon flux of $3.2\times 10^{-5}$
cm$^{-2}$ s$^{-1}$, corresponding to $W_{\rm Fe}=\blad{530}{320}{260}$ eV. If the line were
intrinsic, its energy indicating origin in a strongly ionized medium and the
large equivalent width would be difficult to explain given the very low
luminosity of the source, when the accretion disc is expected to be cold and far
away from the black hole. For example, the Fe K emission is weak in a faint hard
state of the black-hole binary XTE J1118+480, at an Edddington ratio somewhat
higher than that of the present observation (Frontera et al.\ 2001b).

Thus, we have investigated whether the line may be {\it not\/} intrinsic to GX
339--4. In fact, the source lies close to the direction of the Galactic centre,
where the Galactic diffuse emission is strong.  Based on the observations of
Yamauchi \& Koyama (1993), we have estimated the intensity of the Fe 6.7 keV
line in the Galactic diffuse emission at the direction of GX 339--4 is $\approx
3\times 10^{-5}$ cm$^{-2}$ s$^{-1}$ deg$^{-2}$. Given the field of view of the
PCA of $\sim 1$ deg$^2$, the Fe K line seen in our data should be attributed
mostly to the Galactic background.

Similarly strong lines were reported and attributed to GX 339--4 in a number of 
off-state PCA observations (including ours) by Feng et al.\ (2001), but we 
believe their origin is also diffuse. Furthermore, Nowak et al.\ (2002) found 
that the ratio of the line equivalent width to the relative reflection strength 
in GX 339--4 was increasing with the decreasing luminosity, which they claimed 
to represent an argument against simple reprocessing models of those features. 

However, we have checked that the effect seen by Nowak et al.\ (2002) can be  
explained by the relative contribution of the diffuse 6.7 keV line increasing 
with the decreasing luminosity. For the two weakest of their observations, where the $3-9$ keV source flux was by factors of $\sim 13$ and $\sim 40$ higher than in our observation, the total line equivalent width was $W_{\rm Fe}=150$ eV and $W_{\rm Fe}=160$ eV, respectively. The Galactic diffuse emission would then contribute to the observed $W_{\rm Fe}$ at the level of  $\approx 40$ eV and $\approx 13$ eV, respectively. Then, the agreement between the intrinsic Fe K$\alpha$ line equivalent width and the strength of reflection in their data would be significantly improved.

\section{The origin of seed photons and constraints on magnetic field}
\label{s:nth}

Here, we explore some consequences of the possible presence of a non-thermal 
electron tail, found statistically necessary in our uniform-plasma model for the 
fourth observation. A weak non-thermal tail beyond a Maxwellian distribution 
leads to a moderate hardening of the X-ray power law in the Comptonization 
spectrum (as repeated scattering off thermal electrons still dominates at those 
energies) as well as it yields a weak high-energy photon tail at energies $\gg 
kT$ (from scattering off non-thermal electrons), see, e.g., WZ01. On the other 
hand, it can amplify the cyclo-synchrotron emission by a very large factor, due 
to the fact that the unabsorbed synchrotron radiation is emitted just by the 
high-energy tail of the electron distribution, which itself is strongly 
amplified by the tail with respect to the Maxwellian (WZ01). This amplified 
synchrotron emission within the Comptonizing plasma provides then a copious 
source of seed photons. For given magnetic field strength, $B$, and the plasma 
parameters, $kT$ and $\tau$, we can calculate the expected self-Compton 
luminosity, which can exceed the observed $L$ in general.

This provides a way to constrain $B$, with the maximum one corresponding to all
the seed photons being synchrotron (see section 5.3 of WZ01). Since we know that
a cold medium is also present in the plasma vicinity, which medium is also a
copious source of soft blackbody photons, the actual value of $B$ has to be even lower.

First, we consider a hot accretion flow with nearly-virial ions (e.g., 
Abramowicz et al.\ 1995; Narayan \& Yi 1995). Following WZ00, we estimate the 
equipartition field strength within a radius of $20GM/c^2$ (where most of the 
gravitational energy is dissipated) as $B\simeq 1.4\times 10^7$ G. Then, 
following the calculations in WZ01 and using the hybrid electron distribution 
fitted to the data set 4 ($\delta \simeq 0.3$, Section \ref{s:xte}), we obtain 
the synchrotron--self-Compton luminosity of $4.3\times 10^{38}$ erg  s$^{-1}$, 
which exceeds the observed one by a factor of $\sim 40$. The condition that the 
predicted luminosity does not exceed the observed $L$ leads to $B\la 2.2\times 
10^6$ G. Although this is a factor of several below the equipartition, various 
processes dissipating magnetic field can lead to this condition to be satisfied 
in an actual accretion flow. On the other hand, the self-Compton luminosity from 
the purely thermal electron distribution fitted to the data set 4  (Section 
\ref{s:xte}) would be only $1.6\times10^{35}$ erg s$^{-1}$, i.e., negligibly 
small compared to the observed $L$. Thus, the non-thermal tail amplifies thus 
the synchrotron emission by a rather large factor of $\sim 3\times 10^3$. 
Consequently, most of the seed photons for Comptonization have to be emitted by 
a cold medium if the tail is very weak or absent.

We have also considered the configuration of small, active regions above a cold
disc, in which regions dissipation of magnetic field occurs (see WZ00 for
details of the assumed model). Assuming that magnetic field dissipates at 10 per
cent of the Alfv\'en speed in 10 active regions of the size $10^6$ cm, the
magnetic field required to dissipate enough energy to reproduce the observed $L$
would be $B\simeq 10^8$ G. Then, however, the synchrotron--self-Compton
luminosity from the hybrid plasma would be $\sim 20$ times the observed $L$. It
would be $\leq L$ for $B\la 2.1\times 10^7$ G, which field, however, would not
be able to dissipate enough energy to power the observed $L$. We thus conclude
that the non-thermal electron tail fitted to the data rules out the active
regions model for the hard state of GX 339--4. The same conclusion was obtained
by WZ01 for the hard state of Cyg X-1. Also, Z98 found the model of active
regions for the hard state of GX 339--4 to be ruled out based on independent
spectral considerations.

\section{The origin of the optical emission in the hard state}
\label{s:optical}

GX 339--4 in an X-ray hard state was observed  by the {\it Ariel 6}\/ satellite
(Ricketts 1983), which observation was in part simultaneous with optical
observations, which found the source in a very bright state (Motch, Ilovaisky \&
Chevalier 1981, 1982). The simultaneous data from 1981 May 28 are analyzed by
M83. The 1--50 keV X-ray spectrum of Ricketts (1983) is, in fact, very similar
to our \ginga-OSSE spectra. Given our considerations of the origin of the seed
photons for Comptonization in the X-ray regime, it is of great interest to
check whether the optical emission can be due to the cyclo-synchrotron
process in the X-ray emitting plasma. Such origin of the optical emission was
suggested by Fabian et al.\ (1982), but those authors have not performed any
calculations of the actual efficiency of that process.

\begin{figure}
\begin{center}
\noindent\epsfxsize=8.2cm\epsfbox{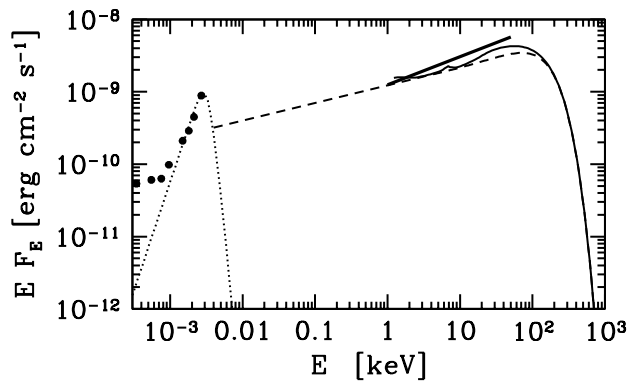}
\end{center}
\caption{The dereddened IR/optical spectrum (circles) of GX 339--4 from 1981 May
24--28 (circles) and the 1--50 keV power-law fit (thick line) to the 1981 May
30--31 observation. The thin solid curve shows
the thermal Comptonization and reflection components of the model to our data
set 1 (from \ginga\/ and OSSE), and the dashed curve shows the Comptonization
component alone, including its extrapolation to low energies. The thin dotted
curve shows a self-absorbed thermal synchrotron spectrum.
}
\label{f:optical}
\end{figure}

Here, we assume that the plasma temperature and optical depth during the 1981
observation were the same as those obtained from our X-ray data and consider in
detail the requirements on other parameters implied by the cyclo-synchrotron
origin of the optical emission. Fig.\ \ref{f:optical} shows the unabsorbed
spectrum of GX 339--4 from the IR to soft \g-rays. The IR/optical data are from
Motch et al.\ (1981), dereddened assuming $E(B-V)=1.2$, and the X-ray spectrum
is from Ricketts (1983).

Fig.\ \ref{f:optical} also shows the \ginga-OSSE spectrum. We see that the
spectra from \ginga-OSSE and {\it Ariel 6\/} are indeed quite similar.
Furthermore, the extrapolation to the optical range of the thermal-Compton
spectrum fitted to the \ginga-OSSE spectrum lies just a factor of several below
the peak of the optical emission. Thus, the optical emission can indeed provide
enough seed photons for Comptonization in the X-ray emitting plasma.

\begin{figure}
\begin{center}
\noindent\epsfxsize=6.8cm\epsfbox{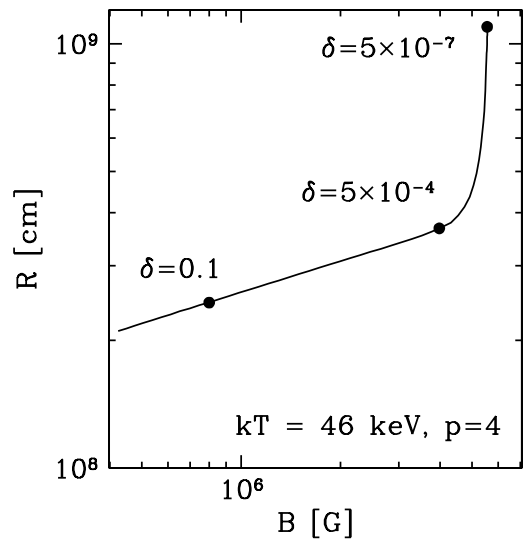}	
\end{center}
\caption{The source size and its magnetic field strength required to explain the
optical spectrum of 1981 May as due to cyclo-synchrotron emission parametrized
by the relative fraction of energy in non-thermal electrons, $\delta$. }
\label{B_R}
\end{figure}

We then assume that the uppermost point in the optical spectrum, $E F_{\rm
E}\simeq 9\times 10^{-10}$ erg cm$^{-2}$ s$^{-1}$ at $E\simeq 2.7$ eV,
corresponds to cyclo-synchrotron emission at the turnover frequency (i.e., where
the plasma becomes optically thin to synchrotron self-absorption). The
corresponding spectrum is shown by the dotted curve in Fig.\ \ref{f:optical}. In
the purely thermal case and with the plasma parameters as in Table \ref{t:fits},
fit 1, the required magnetic field strength and the size of the plasma (assuming
spherical geometry, and calculated as in WZ00) are $B\sim 10^7$ G and $R\sim
10^9$ cm, respectively. (These values are each an order of magnitute higher than
those given by Fabian et al.\ 1982.) The required source size is rather large,
$\sim 10^3 GM/c^2$. It can be reduced if either the magnetic field were larger
(corresponding to the synchrotron turnover peak at both higher photon energy and
flux, which is allowed by the data), $kT$ were higher (as its value during the
1981 observation is unknown), or the electron distribution contained a
non-thermal tail. The last possibility is, in fact, suggested by our data
(Section \ref{s:xte}), and we consider it in detail. We assume the electron
power law index of $p=4$, as before, and calculate the required values of $B$
and $R$ as a function of the fraction of energy in the non-thermal electrons,
$\delta$. Results are shown in Fig.\ \ref{B_R}. We see that $R$ is reduced by a
factor of several already for $\delta\sim 5\times 10^{-4}$, and for $\delta\sim
0.1$, both $R$ and $B$ are reduced by an order of magnitude. The field expected
in a hot accretion disc flow in equipartition with nearly-virial ions would be
$B\sim 10^6$ G and $\sim 10^5$ G at at $R\sim 10^8$ and $10^9$ cm, respectively.
Thus, $R\sim 10^2 GM/c^2 \sim 10^8$ cm, the equipartition $B\sim 10^6$ G, and
$\delta \sim 0.1$ represent our preferred set of the plasma parameters of the
flow explaining the optical emission.

Interestingly, M83 found an anticorrelation between the optical emission and the
1--13 keV X-rays. An anticorrelation can appear, e.g., if the cyclo-sychrotron
emission is variable (due to variability of some plasma parameters) but the
total luminosity is approximately constant. The spectral variability would then
show a pivot between the optical and X-ray ranges, which naturally leads to an
anticorrelation between the respective emission. Furthermore, M83 found that the
optical emission leads the X-rays by $2.8\pm 1.6$ s. At $kT= 46$ keV (Table
\ref{t:fits}), it takes $\sim 30$ scatterings to upscatter a 2.7 eV photon to
10 keV. At $\tau\sim 3$, the source size required for the time lag to be due to
the delay during Compton upscattering is then $\sim 10^9$ cm, still a plausible
value. M83 also found a QPO at $\sim 20$ s in their data, which, interestingly,
corresponds to the Keplerian frequency at a radius of $2\times 10^9$ cm (at
$M=10 \msun$). A somewhat lower QPO period, $\sim 3$ s, was found in the data
set 4 (R01). The time scale of flares of 20 ms found by M83
can then correspond to the light travel time across the source. A remaining
unresolved issue would be then the lack of an anticorrelation found by M83
between the optical photons and those in the 13--20 keV range. Possibly, that
energy range is dominated by Compton reflection showing a different timing
behaviour than the primary continuum.

On the other hand, Fabian et al.\ (1982) proposed that the anticorrelation is
due to the X-rays below 13 keV being due to emission of colder clouds
surrounding the hot plasma (emitting both the optical emission and X-rays above
13 keV). However, present fits to X-ray spectra of GX 339--4 as well as other
black-hole binaries in the hard state tend to rule out this possibility,
indicating that the entire X-ray emission is dominated by a single power law
(e.g., U94; Z98, Gierli\'nski et al.\ 1997).

We note that similar optical/X-ray behaviour has been observed from the 
black-hole binary XTE J1118+480 (Kanbach et al.\ 2001). Optical emission from 
that object is very strong, comparable to X-rays, similar to the case of GX 
339--4. Also, a similar anti-correlation between optical and X-ray flux was 
observed, with the optical minimum leading the X-rays maximum by $\sim 2$ s. 
Furthermore, the optical emission has a narrower auto-correlation function than 
that of X-rays, which implies that the primary variability is in the optical 
band. It is then possible that the optical emission is cyclo-synchrotron, which 
is then Comptonized up to the X-ray range. The anticorrelation can be explained 
as in GX 339--4, by spectral pivoting between the optical and X-ray bands.

The site of the hot plasma could be in principle an outflow, possibly a jet, 
which presence is indicated by radio observations in the hard state (e.g., 
Fender et al.\ 1997, 1999; Corbel et al.\ 2000). In fact, double-peaked emission 
lines have been found in the soft state of GX 339--4 by Soria et al.\ (1999), 
whereas they observed broad, single-peaked profiles in the hard state. They 
attributed the difference in the line profiles to the origin of the line in the 
outflow in the hard state and in the disc in the soft state. Also Wu et al.\ 
(2001) reached a similar conclusion. On the other hand, there is no natural 
interpretation to the frequency of the optical QPO in the outflow model. A way 
to test the origin of the optical radiation would be to measure its 
polarization, which would be non-zero in an outflow with a directed magnetic 
field. 

While the optical spectrum may in principle be either quasi-thermal and produced 
in an accretion flow or purely non-thermal and produced in a jet, our fits to 
the \g-ray high-energy cutoff confirm that the distribution of electrons 
producing the main X/$\gamma$-ray continuum in the hard state has a 
predominantly thermal character (even if it does originate in an outflow, as, 
e.g., in the coronal outflow model of Beloborodov 1999a). This represents a 
strong argument against the model of the X/$\gamma$-ray continuum of black-hole 
binaries as being predominantly non-thermal (Markoff, Falcke \& Fender 2001). In 
that model, the predicted position of the cutoff in the photon spectrum is very 
sensitive to the magnetic field strength and details of acceleration and 
cooling. Markoff et al.\ (2001) deal with this problem by fine-tuning the 
high-energy cutoff in the distribution of the non-thermal electrons. However, 
virtually all black-hole binaries in the hard state have remarkably uniform 
high-energy cutoffs (e.g., Grove et al.\ 1998; Z98; Gierli\'nski et al.\ 1997), 
which thus represents a severe problem for the non-thermal model. On the other 
hand, this uniformity is accounted for in the thermal (or quasi-thermal) model, 
in which the Compton cooling (e.g., Zdziarski 1998) and/or $e^\pm$ pair 
production (e.g., Malzac, Beloborodov \& Poutanen 2001) naturally constrain the 
electron temperature to $\sim 50$--100 keV. 

\section{X-ray variability}
\label{s:variability}

The fractional variability of GX 339--4 in X-rays was investigated by, e.g., 
Maejima et al.\ (1984), Nowak et al.\ (1999), Lin et al.\ (2000) and R01 and was 
found to be, in general, energy-dependent. Here, we investigate the 
energy-dependence of the X-ray variability in the \xte\/ data of the set 4, 
which has the best statistics of the data sets studied by us. We use all layers 
in all available PCA detectors (PCUs 0--4) in order to obtain the best possible 
photon statistics.

\begin{figure}
\begin{center}
\noindent\epsfxsize=6.8cm\epsfbox{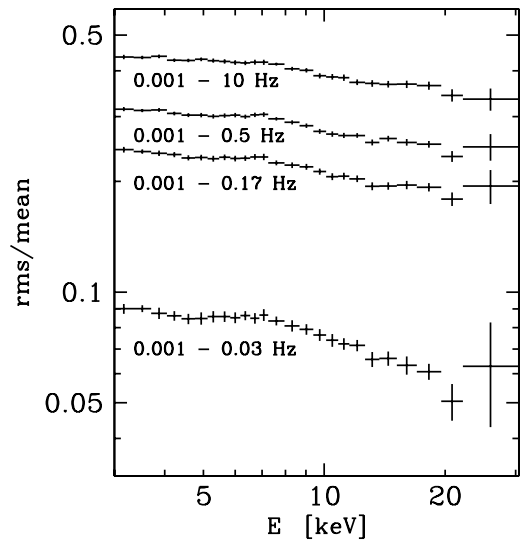}	
\end{center}
\caption{The fractional rms X-ray variability of GX 339--4 in the PCA data of
the set 4 for four ranges of frequency.
}
\label{f:var}
\end{figure}

We construct light curves in a number of energy bands with different bin size, 
$\Delta T$. For each lightcurve, we integrate the power density spectrum in the 
frequency range from 0.001 Hz up to the upper frequency of $1/(2\Delta T)$, from 
which we obtain the (Poisson-corrected) rms variability divided by the mean 
count rate. The latter has been corrected for the instrumental background, which 
is strong, especially at $E\ga 20$ keV. The results are presented in Fig.\ 
\ref{f:var}.

A common feature of all the shown dependencies is a decrease of the rms/mean
starting at $E\sim 7$ keV. This agrees with a general trend of the fractional
rms decreasing with increasing X-ray energy in other black-hole binaries in the
hard state (Revnivtsev et al.\ 2000). In our case, the ratio of the rms at $E\gg
7$ keV to that at $E\ll 7$ keV increases with the increasing upper frequency.

R01 have studied a number of PCA data for GX 339--4 including our set 4 and 
found that the reflection component is much more variable at $\la 0.1$ Hz than 
at $\sim 10$ Hz. This effect will result in a reduction of the rms at $E\ga 10$ 
keV. However, we have checked that the dependencies of rms on $E$ in Fig.\ 
\ref{f:var} cannot be explained just by a superposition of different values of 
the rms for the power law and for reflection. Also, the ratio of the rms/mean at 
high energies to that at low energies would decrease with the increasing upper 
frequency whereas we observe the opposite behaviour.

It is thus more likely that we observe a change of the variability pattern in
the Comptonization continuum itself. We propose that it could take place in a
similar scenario to that considered in Section \ref{s:optical} when analyzing
the optical and X-ray variability. Namely, if the total Comptonization
luminosity remains constant but the flux in seed photons varied, this would
result in a pivoting of the spectrum. The pivot energy would in general depend
on the spectral index and the temperatures of the electrons and the seed
photons. Because a part of the spectrum closer to the pivot energy would be less
variable than that further away, this scenario can qualitatively reproduce the
observed rms/mean for the pivot at $E\ga 20$ keV. The dependence on the upper
frequency can be explained if the pivoting occured mostly on long time scales.
We note that a similar pivoting is indeed observed in Cyg X-1 on very long
time scales (Zdziarski et al.\ 2002).

We note that the pivoting energy required by the above interpretation is
different from that needed for explaining the optical/X-ray anticorrelation by
varying cyclo-synchrotron seed photons in Section \ref{s:optical}. For that, the
pivot energy has to be obviously between the optical and X-ray ranges. Thus, the
two interpretations are mutually exclusive. We have also calculated that the
pivot at $E\ga 20$ keV implies $k T_{\rm seed} \ga 0.1$ keV, which rules out the
cyclo-synchrotron origin of the seed photons in our data set 4. We point out,
however, that the two variability patterns were observed in two different
observations of GX 339--4 (separated by 15 years), during which the state of the
source could be different. Also, the cause of the dependence of the rms on $E$
shown in Fig.\ \ref{f:var} could be different than due to pivoting. It can
possibly be explained, e.g., by models of magnetic avalanches (e.g., Poutanen \&
Fabian 1999), which predict complicated patterns of variability of flares on
different time scales.

\section{Conclusions}
\label{s:summary}

We have analyzed four broad band X/$\gamma$-ray spectra of GX 339--4 in the hard 
state and found that they are well fitted by Comptonization being predominantly 
thermal, Compton reflection and fluorescent Fe K emission with the strength 
compatible with that of the reflection. This provides a further confirmation of 
the dominance of those physical processes in this object, and black-hole 
binaries in the hard state in general. It also argues against the non-thermal 
model for the origin of the X/$\gamma$-ray continuum of black-hole binaries by 
Markoff et al.\ (2001). We also study time variability in our best X-ray data 
set, and find a decline of the fractional rms with increasing photon energy, 
again consistent with previous findings for this source.

Our fifth observation was in an off state, with the X-ray flux two orders of
magntitude below that in the hard state. Our major finding here is that the very
strong Fe K line present in our data (as well as in other reported
proportional-counter off-state observations of this source) is {\it not\/}
intrinsic to GX 339--4, but is due to the strong diffuse 6.7 keV emission in the
Galactic Center region.

We also obtain a number of other results, which, however, are less unambiguous
and require future further studies. This is due to either the limitations of the
available data or insufficient theoretical understanding at the present time.

In the off state, we find a spectral hardening at $\sim 7$ keV, which we find to 
be incompatible with thermal Comptonization but, rather, compatible with the 
presence of a bremsstrahlung spectral component. Such a component is in 
agreement with theoretical predictions, but its unambiguous detection is 
hampered by the low signal-to-noise ratio and lack of sufficient coverage at 
high energies.

In the hard state, we find a statistically significant departure from the pure
thermal Comptonization spectrum at high energies in our best spectrum. We
interpret it as due to the presence of a weak high-energy tail in the electron
distribution. However, we cannot rule out an alternative explanation of the
effect of the plasma parameters varying in space and/or time.

The possible presence of a high-energy electron tail has implications for
the strength of the magnetic field in the source, from the consideration of
Compton scattering of the cyclo-synchrotron emission. The magnetic field
strength is constrained to values somewhat below equipartition (with hot ions)
in the case of a hot accretion disc.

Interestingly, the four hard-state spectra have very similar intrinsic X-ray 
slope, $\Gamma\simeq 1.75$.  On the other hand, the high-energy cutoff energy 
decreases with increasing luminosity, and, correspondingly, the fitted Thomson 
optical depth of the plasma increases with luminosity. The significance of the 
correlations is strongly increased if we allow for the presence of a high-energy 
electron tail in our model. Such a correlation is in agreement with theoretical 
predictions for hot accretion flows. Still, it is based only on the four 
available spectra, and studies of more broad-band spectra with similar X-ray 
slope, $\Gamma$, are highly desirable.

We also study the origin of the strong optical emission found in the past to be 
associated with the X-ray bright hard state. We find the reported optical 
emission can be due to the cyclo-sychrotron process, and the emission also 
provides seed photons for Comptonization. The presence of a high-energy electron 
tail alleviates constraints on the emission region, allowing it to be smaller 
and with a weaker field. The reported optical/X-ray anticorrelation can be due 
to the synchrotron emission being variable at an approximately constant total 
luminosity, which leads to a spectral pivot between the optical and X-ray bands.

On the other hand, we find the dependence of the rms variability on energy can
be also explained by variable seed photons, but the pivot point is then at $\ga
20$ keV, i.e., in conflict with the previous model. Thus, a comprehensive model
of both of these variability patterns requires more complications, probably more
source components.

\section*{ACKNOWLEDGMENTS}
This research has been supported in part by the Foundation for Polish Science
and KBN grants 5P03D00121, 5P03D00821 and 2P03D00514. We are very grateful to Mike Revnivtsev for his help with the variability analysis and the suggestion
to consider the diffuse Fe K emission. We also thank Bo\.zena Czerny, Chris
Done, Piotr Lubi\'nski, Joanna Miko{\l}ajewska, Marek Sikora and Piotr \.Zycki
for valuable and inspiring discussions.

\bsp

\label{lastpage}


\begin{thebibliography}{}

\bibitem{a95}
Abramowicz M. A., Chen X., Kato S., Lasota J.-P., Regev O., 1995, ApJ, 438, L37

\bibitem{ae82}
Anders E., Ebihara M., 1982, Geochim.\ Cosmochim.\ Acta, 46, 2363

\bibitem{a96}
Arnaud K. A., 1996, in Jacoby G. H., Barnes J., eds., Astronomical Data Analysis
Software and Systems V, ASP Conf. Series Vol.\ 101, San Francisco, p.\ 17

\bibitem{a98}
Asai K., Dotani T., Hoshi R., Tanaka Y., Robinson C. R., Terada K., 1998, PASJ,
50, 611

\bibitem{2001MNRAS.327...10B} 
Ballantyne D.~R., Ross R.~R., Fabian A.~C., 2001, MNRAS,  327, 10 

\bibitem{b99a}
Beloborodov A.  M., 1999a, ApJ, 510, L123

\bibitem{b99b}
Beloborodov A. M., 1999b, in Poutanen J., Svensson, R. eds., High
Energy Processes in Accreting Black Holes, ASP Conf.\ Ser.\ Vol.\ 161,
San Francisco, p.\ 295

\bibitem{br92}
Bevington P. R., Robinson K. D., 1992, Data Reduction and Error Analysis
for the Physical Sciences, 2nd Ed., New York, McGraw-Hill

\bibitem{b98}
B\"ottcher M., Liang E. P., Smith I. A., 1998, A\&A, 339, 87

\bibitem{b93}
Bouchet L. et al., ApJ, 1993, 407, 739

\bibitem{c00}
Corbel S., Fender R. P., Tzioumis A. K., Nowak M., McIntyre V., Durouchoux P.,
Sood R., 2000, A\&A, 359, 251

\bibitem{d01}
Di Salvo T., Done C., \.Zycki P. T., Burderi L., Robba N. R., 2001, ApJ, 547,
1024

\bibitem{d92}
Done C., Mulchaey J. S., Mushotzky R. F., Arnaud K. A., 1992, ApJ, 395, 275

\bibitem{d98}
Dove, J. B., Wilms, J., Nowak, M. A., Vaughan, B. A., Begelman, M. C.,
1998, MNRAS, 298, 729

\bibitem{e97}
Esin A. A., McClintock J. E., Narayan R., 1997, ApJ, 489, 865

\bibitem{f82}
Fabian A. C., Guilbert P. W., Motch C., Ricketts M., Ilovaisky S. A., Chevalier
C., 1982, A\&A, 111, L9

\bibitem{f89}
Fabian A. C., Rees M. J., Stella L., White N. E., 1989, MNRAS, 238, 729

\bibitem{f97}
Fender R. P., Spencer E. R., Newell S. J., Tzioumis A. K., 1997, MNRAS, 286, L29

\bibitem{f99}
Fender R. P. et al., 1999, ApJ, 519, L165

\bibitem{f01}
Feng Y. X., Zhang S. N., Sun X., Durochoux P., Chen W., Cui W., 2001, 553, 394

\bibitem{f01a}
Frontera F. et al., 2001a, ApJ, 546, 1027

\bibitem{f01b}
Frontera F. et al., 2001b, ApJ, 561, 1006

\bibitem{gf91}
George I. M., Fabian A. C., 1991, MNRAS, 249, 352

\bibitem{gz97}
Gierli\'nski M., Zdziarski A. A, Done C., Johnson W. N., Ebisawa K., Ueda Y.,
Haardt F., Phlips B. F., 1997, MNRAS, 288, 958

\bibitem{g99}
Gierli\'nski M., Zdziarski A. A., Poutanen J., Coppi P. S., Ebisawa K., Johnson
N. W., 1999, MNRAS, 309, 496

\bibitem{g00}
Gilfanov M., Churazov E., Revnivtsev M., 2000, in G. Zhao et al., eds., Proc.\
5th CAS/MPG Workshop on High Energy Astrophysics, Beijing,
Sci.\ Techn.\ Press, p.\ 114 (astro-ph/0002415)

\bibitem{g95}
Grabelsky D. A. et al., 1995, ApJ, 441, 800

\bibitem{g79}
Grindlay J. E., 1979, ApJ, 232, L33

\bibitem{g98}
Grove J. E., Johnson W. N., Kroeger R. A., McNaron-Brown K., Skibo J. G., 1998, ApJ, 500, 899

\bibitem{k01}
Kanbach G., Straubmeier C., Spruit H. C., Belloni T., 2001, Nat, 414, 180

\bibitem{k00}
Kong A. K. H., Kuulkers E., Charles P. A., Homer L., 2000, MNRAS, 312, L49

\bibitem{l00}
Lin D., Smith I. A., B\"ottcher M., Liang E. P., 2000, ApJ, 531, 963

\bibitem{m84}
Maejima Y., Makishima K., Matsuoka M., Ogawara Y., Oda M., Tawara Y., Doi K.,
1984, ApJ, 285, 712

\bibitem{mz95}
Magdziarz P., Zdziarski A. A., 1995, MNRAS, 273, 837

\bibitem{m86}
Makishima K. et al., 1986, ApJ, 308, 635

\bibitem{mbp01}
Malzac J., Beloborodov A. M., Poutanen J., 2001, MNRAS, 326, 417

\bibitem{m73}
Markert T. H., Canizares C. R., Clark G. W., Lewin W. H. G., Schnopper H. W.,
Sprott G. F., 1973, ApJ, 184, L67

\bibitem{m01}
Markoff S., Falcke H., Fender R., 2001, A\&A, 372, L25
\bibitem{mc00}
McConnell M. L. et al., 2000, ApJ, 543, 928

\bibitem{mc02}
McConnell M. L. et al., 2002, ApJ, 572, 984

\bibitem{m81}
Motch C, Ilovaisky S. A., Chevalier C., 1981, IAUC 3610

\bibitem{m82}
Motch C, Ilovaisky S. A., Chevalier C., 1982, A\&A, 109, L1

\bibitem{m83}
Motch C., Ricketts M. J., Page C. G., Ilovaisky S. A., Chevalier C., 1983, A\&A,
119, 171 (M83)

\bibitem{np94}
Nandra K., Pounds K., 1994, MNRAS, 268, 405

\bibitem{ny95}
Narayan, R., \& Yi, I. 1995, ApJ, 452, 710

\bibitem{n99}
Nowak M. A., Wilms J., Dove J. B., 1999, ApJ, 517, 355

\bibitem{n02}
Nowak M. A., Wilms J., Dove J. B., 2002, MNRAS, 332, 856

\bibitem{ps96}
Poutanen J., Svensson R., 1996, ApJ, 470, 249

\bibitem{pf99}
Poutanen J., Fabian A. C., 1999, MNRAS, 306, L31

\bibitem{p92}
Press W. H., Teukolsky S. A., Vetterling W. T., Flannery B. P., 1992, Numerical
Recipes. Cambridge Univ.\ Press, Cambridge

\bibitem{r01}
Revnivtsev M., Gilfanov M., Churazov E., 2001, A\&A, 380, 520 (R01)

\bibitem{f00}
Revnivtsev M., Borozdin K. N., Priedhorsky W. C., Vikhlinin A., 2000, ApJ, 530,
955

\bibitem{r83}
Ricketts M. J., 1983, A\&A, 118, L3

\bibitem{sl99}
Smith I. A., Liang E. P., 1999, ApJ 519, 771

\bibitem{s99}
Smith I. A. et al., 1999, ApJ, 519, 762

%\bibitem{}
%Smith I. A., Filippenko A. V., Leonard D. C., 1999, ApJ, 519, 779

\bibitem{swj99}
Soria R., Wu K., Johnston H. M., 1999, MNRAS, 310, 71

\bibitem{sr00}
Sunyaev R. A., Revnivtsev M., 2000, A\&A, 358, 617

\bibitem{st79}
Sunyaev R. A., 	Tr\"{u}mper J., 1979, Nat, 279, 506

\bibitem{t94}
Titarchuk L., 1994, ApJ, 434 313

\bibitem{t98}
Trudolyubov S. et al., 1998, A\&A, 334, 895

\bibitem{u94}
Ueda Y., Ebisawa K. Done C., 1994, PASJ, 46, 107 (U94)

\bibitem{w00}
Wardzi\'nski G., Zdziarski, A. A., 2000, MNRAS, 314, 183 (WZ00)

\bibitem{w01}
Wardzi\'nski G., Zdziarski, A. A., 2001, MNRAS, 325,963; Erratum: 2001, MNRAS,
327, 351 (WZ01)

\bibitem{w99}
Wilms J., Nowak M. A., Dove J. B., Fender R. P., Di Matteo T., 1999, ApJ, 522,
460

\bibitem{wd01}
Wilson C. D., Done C., 2001, MNRAS, 325, 167

\bibitem{wu01}
Wu K., Soria R., Hunstead R. W., Johnston H. M., 2001, MNRAS, 320, 177

\bibitem{yk93}
Yamauchi S., Koyama K., 1993, ApJ, 404, 620

\bibitem{z98}
Zdziarski A. A., 1998, MNRAS, 296, L51

\bibitem{z00}
Zdziarski A. A., 2000, in Martens P. C. H., Tsuruta S., Weber M. A., eds.,
Highly Energetic Physical Processes and Mechanisms for Emission
from Astrophysical Plasmas, IAU Symp. 195, San Francisco, ASP, p.\ 153
(astro-ph/0001078)

\bibitem{Zdz95}
Zdziarski A. A., Johnson W. N., Done C., Smith D., McNaron-Brown K., 
1995, ApJ, 438, L63

\bibitem{zpm98}
Zdziarski A. A., Poutanen J., Miko{\l}ajewska J., Gierli\'nski M., Ebisawa
K., Johnson W. N., 1998, MNRAS, 301, 435 (Z98)

\bibitem{z99}
Zdziarski A.  A., Lubi\'nski P., Smith D.  A., 1999, MNRAS, 303, L11

\bibitem{z02}
Zdziarski A.  A., Poutanen J., Paciesas W. S., Wen L., 2002, ApJ, 578, in press
(astro-ph/0204153)

\bibitem{zc94}
\.Zycki P. T., Czerny B., 1994, MNRAS, 266, 653

\end{thebibliography}
\end{document}